# Advancing low-field MRI with a universal denoising imaging transformer: Towards fast and high-quality imaging

Zheren Zhu[1], Azaan Rehman[2], Xiaozhi Cao[3,4],
Congyu Liao[3,4], Yoo Jin Lee[1], Michael Ohliger[1], Hui Xue[2]*, Yang Yang[1]*

[1] Department of Radiology and Biomedical Imaging, University of California, San Francisco, CA, United States
[2] National Heart, Lung, and Blood Institute, Bethesda, MD, United States
[3] Department of Radiology, Stanford University, Stanford, CA, United States
[4] Department of Electrical Engineering, Stanford University, Stanford, CA, United States

**Word count:** 5813

**\*Correspondence**

Hui Xue, PhD
National Heart, Lung, and Blood Institute
National Institutes of Health
10 Center Drive, Bethesda, MD, United States
Email: hui.xue@nih.gov

Yang Yang, PhD
Department of Radiology and Biomedical Imaging
University of California
San Francisco, CA, United States
Email: yang.yang4@ucsf.edu

## Abstract

Recent developments in low-field (LF) magnetic resonance imaging (MRI) systems present remarkable opportunities for affordable and widespread MRI access. A robust denoising method to overcome the intrinsic low signal-noise-ratio (SNR) barrier is critical to the success of LF MRI. However, current data-driven MRI denoising methods predominantly handle magnitude images and rely on customized models with constrained data diversity and quantity, which exhibit limited generalizability in clinical applications across diverse MRI systems, pulse sequences, and organs. In this study, we present ImT-MRD: a complex-valued imaging transformer trained on a vast number of clinical MRI scans aiming at universal MR denoising at LF systems. Compared with averaging multiple-repeated scans for higher image SNR, the model obtains better image quality from fewer repetitions, demonstrating its capability for accelerating scans under various clinical settings. Moreover, with its complex-valued image input, the model can denoise intermediate results before advanced post-processing and prepare high-quality data for further MRI research. By delivering universal and accurate denoising across clinical and research tasks, our model holds great promise to expedite the evolution of LF MRI for accessible and equal biomedical applications.

## Introduction

Magnetic resonance imaging (MRI) is a fundamental cornerstone in contemporary medical diagnostics, providing a powerful and versatile tool for precise clinical diagnosis. Over the past three decades, MRI has predominantly been performed at high magnetic field strength, namely 1.5T and 3T. Higher field strength leads to a proportionally increased signal-to-noise ratio (SNR), enabling higher spatial resolution or faster imaging. On the other hand, high-field-strength superconducting magnets require large amounts of energy, cooling capacities, more complicated shimming systems, and shielded rooms, making these scanners extremely costly to purchase, install, maintain, and operate. Consequently, their availability in clinical settings is extremely insufficient and unequally distributed worldwide. For instance, while the United States averages 40 MRI scanners per million people, Africa has only 0.7[1]. Even in high‑income countries, MRI is hardly accessible in rural areas[2].

In recent years, low-field[3] MR scanners, including 0.55T models, have undergone intensive development and become available, offering wide-spreading potential and clinical value for their superior cost-efficiency and accessibility. However, their intrinsically decreased SNR often results in compromised image quality or, most commonly, requires longer scan duration (e.g., multiple-repeated scanning) to restore lost SNR[4]. This significantly diminishes clinical efficiency and reduces revenue, in addition to poor patient experience.

Deep learning models have shown great promise for denoising MR images, enabling faster scans and better image quality due to their effectiveness in detail preservation and low-SNR recovery compared to conventional post-processing-based denoising methods[5]. However, one of the main limitations of many current deep learning MRI denoising models[6–9] is their task-specific nature and very poor generalization. In MRI, the signal distribution of raw k-space data can vary significantly among patients, imaging sequences, and MRI systems, reflecting the considerable diversity of signal and noise distribution in MRI images. These models are typically developed and trained for a specific denoising task or targeted noise level, and their performance can degrade significantly when applied to a different clinical setting. On the other hand, many current denoising models[10,11,8,9] are centered around recovering magnitude signals of MRI images. These are usually the last-step results throughout the entire pipeline of MRI image reconstruction and possess the least MR-specific information. This will not only cause the model to miss rich details during the training but also give the least flexibility to apply denoising in advanced MRI techniques that require additional post-processing in complex value format.

In this study, we aim to advance low-field MRI for speed and quality through a universal deep-learning pathway. We propose ImT-MRD, a universal MRI denoising imaging transformer model trained on large-scale and diverse data from over 4500 raw clinical MR scans covering various organs, pulse sequences, and MRI systems. We evaluate our model's universal denoising performance over comprehensive internal and external validation sets covering various MR imaging settings while, at the same time, including comparisons with the conventional non-data-driven BM3D denoising method, the state-of-the-art (SOTA) deep-learning practical blind image denoising model, and a specialist model trained on the dataset from one specific task. To assess our model's capability of accelerating low-field MRI scans and improving image quality, we compare the averaged images from multi-repeated scans at 0.3T and 0.55T, which is a common technique to raise SNR in low-field systems, with the results denoised from only single-repetition images using our model. An additional experiment on denoising intermediate results with an advanced Magnetic Resonance Fingerprinting technique at 0.55T is also conducted to demonstrate its adaptivity. These experiments highlight the model's potential as a powerful tool for universal low-field MRI denoising and a pathway for accessible MRI.

## Results

**An overview of the denoising model:**

Our model aims to offer a universal and adaptive solution for low-field MRI denoising that generalizes to different MRI protocols and processes complex MR images. A crucial aspect of training such a model is the capacity to accommodate a wide range of variations in practical imaging settings. To address this challenge, we collected a large dataset of 8583 complex-valued high-SNR MRI images from a clinical-based 3T scanner for training, each comprising 30 slices, from 4584 raw MRI scans, covering 29 anatomic regions, 792 different pulse sequence configurations, and 925 patients. This large-scale and diverse dataset allows our model to learn a broad spectrum of anatomy, contrast, resolution, and noise distribution. Figure 1b presents an overview of the training data composition.

At the same time, the complex-valued and stacked image representation enable our model to learn rich, MRI-specified information, encompassing 3D representations through slice-to-slice connectivity and phase details. While complex-valued images provide advantages akin to signal-like features, they also exhibit significant variations in data distribution due to the wide diversity in imaging settings. To address this issue, we introduced a signal-power-based pre-network normalization technique, termed PowerNorm, which applies a factor to scale the original image intensity based on mean image signal power to manage the substantial dataset-wide image intensity variance. Additionally, the factor will be scaled back to the post-network results to retrieve the scaling of the original images.

Another consideration is the synthesis of low-SNR noisy variants for training. To closely mimic practical noise and maintain alignment with real low-field low-SNR imaging, we incorporated geometry (g)-factor[12] maps (gmap) in noise synthesis. Inherent to MR image reconstruction, the noise amplification is spatially variant and depends on the specific geometry of the radiofrequency coil array and is therefore characterized by the g-factor[13]. Adding noise following the gmap distribution allows more realistic clean-noisy training pairs. Considering these challenges, we introduce the pipeline in Figure 1a.

**Internal validation on synthetic dataset**

We evaluated our model through one internal validation with synthetic noise and two external validations with real-world clinical data. For the internal validation, we used 3T high-SNR images plus synthetic noise to simulate low-field low-SNR clinical scans. Here, we included 150 clinical scans from 102 patients, covering various anatomic regions and pulse sequences. The original high-SNR images were used as clean ground truths, while synthetic noise was added to the clean image as the noisy ones. The noise sigma was generated to levels 1, 2, 4, and 6, corresponding to the relative SNR around 32, 26, 20, and 16.48. We compared the denoising performance of our model to the widely used non-learning-based BM3D[14] denoising method with additional noise estimation handling diverse noise intensity and a SOTA practical natural image blind denoising model Swin-Conv-UNet[15] (SCUNet).

Figure 2a shows examples of different protocols at noise level 4. Our model yielded the overall best performance with boosted SNR, consistent noise removal discarding the ununiform noise distribution, and the best preservation of details, while the adjusted BM3D method showed the least improvements in SNR with some detail loss, and the SCUNet model led to severe over-smoothness. Figure 2b shows the peak signal-to-noise ratio (PSNR), structural similarity index measure (SSIM), and normalized root mean square error (NRMSE) scores over the entire validation set of 150 samples at noise level 4. Our model achieved scores of 39.76±3.02, 0.97±0.02, and 0.07±0.01, in average PSNR, SSIM, and NRMSE, which were huge improvements regarding the input noisy images with scores of 32.67±3.38, 0.80±0.09, and 0.16±0.03. Compared to other denoise methods: 35.76±3.35, 0.87±0.08, and 0.11±0.03 for the adjusted BM3D; 37.42±3.64, 0.92±0.06, and 0.09±0.03 for the SOTA model, our model yielded a higher overall PSNR score with higher lower limits, which showed a better and consistent performance in noise removal. Moreover, it provided a better and narrower distribution in SSIM and NRMSE scores, reflecting the robustness of our model across different scenarios.

We further assessed the performance under different noise levels. Figure 2c shows the average PSNR, SSIM, and NRMSE scores under various noise levels from 1 to 6 over the validation set. As shown in Supplementary Figure 1, our model showed excellent consistency in noise removal as well as detail maintenance with the modestly decreasing PSNR and SSIM scores from their peaks when the input images degraded, which demonstrated its capability as a universal denoising model in handling diverse images of different SNRs. On the other hand, the adjusted BM3D denoising results experienced an SNR drop with a loss of details as the noise level went up, and the SOTA blind denoising model exhibited severe over-smoothness. The SOTA model also showed poor consistency when handling different SNRs, especially in high SNR scenarios. These could be attributed to the SOTA model's training on narrow noise level ranges and the fact that practical natural images are more distinct in subjects with higher contrasts.

**External validations on clinical data**

For external validation, we included two datasets (one for routine clinical scans at 0.55T and one for task-specific brain scans at 0.3T) containing multi-repetition raw clinical scan data to test the denoising performance and accelerate potential in real-world scans. For both datasets, the multi-repetition averaged images were used as high-SNR references, which were the results of the conventional scan method at low-field systems, and this substantially extended the scan duration, while the single-repetition ones were used as the low-SNR input before denoising to assess the performance of accelerated scan with denoising.

The first validation dataset includes 150 routine clinical scans from a 0.55T scanner (Siemens MAGNETOM Free. Max). The dataset includes diverse pulse sequences and 13 different anatomic regions (Figure 4a). Each scan contains 4 or 5 times of repetitions without fixed SNR levels but guarantees higher SNR in images after averaging. Figure 3a shows the distribution of the reference PSNR, SSIM, and NRMSE scores over all 150 scan cases, and Figure 3b shows three visualized examples. Our model achieved $36.80\pm3.61$, $0.95\pm0.03$, and $0.11\pm0.03$ in average PSNR, SSIM, and NRMSE, indicating very close results to the multi-repetition averaged images. It also shows better results over the other two denoise methods with scores of $35.98\pm3.67$, $0.91\pm0.04$, and $0.12\pm0.04$ for adjusted BM3D and $36.39\pm3.41$, $0.92\pm0.04$, and $0.11\pm0.04$ for the SOTA.

Beyond quantitative evaluation, two board-certified radiologists assessed the image quality on noise intensity, image sharpness, image details, and overall quality scores. A qualitative analysis was performed on this validation set by two radiologists among the single-repetition noisy inputs, our model's denoising results and the multi-repetition averaged ones. Figure 4b demonstrates that our model not only yielded image quality improvements over single repetitions in all criteria but can also give visually comparable or superior images with less smoothness, better detail preservation, and excellent noise reduction compared to the averaged multi-repetitions.

The second external validation set is the M4Raw[6] test set, an open-sourced task-specific raw image set containing multiple duplications. It comprises three sub-categories (contrasts) of 25 T1, 25 T2, and 25 FLAIR brain imaging cases with fixed scan parameters from a 0.3T permanent-magnet scanner (XGY OPER-0.3T). This validation included a proposed NAFNet[6,16] (Nonlinear Activation Free Network for Image Restoration) specialist model trained with the M4Raw training set (128 T1, 128 T2, and 128 FLAIR brain imaging cases) as a baseline model for comparison. Figure 5a demonstrates examples for each category with category-wise average PSNR and SSIM scores. Figure 5b shows the distribution of reference quantitative scores over the entire dataset across three contrast categories. Overall, the dataset-wise and category-wise reference scores gave similar values among the three deep-learning denoising methods due to the relatively uniform SNR distribution and mild SNR level for all three categories of this dataset according to its dataset quality assessment. The SCUNet output the lowest visual performance with blurred details; meanwhile, both the NAFNet specialist model and our model gave visually preferable results that were close to the multi-repetition averaged references with effective noise removal and detail preservation, indicating the performance of our general model can be comparable to that of a specialist model under its specific task setting.

**Performance in accelerating clinical scans**

To further evaluate the acceleration capabilities of our DL-denoising model in clinical trials, we performed one scan on Lumbar Spine in a 0.55T sagittal STIR sequence[17] with seven repetitions. In this case, we used the average of 7 single-repetitions as a non-accelerated reference, and the averages of 1 to 5 repetitions were used for denoising, representing acceleration rates of 7, 3.5, 2.33, 1.75, and 1.4. As depicted in Figure 6 and Supplementary Figure 2, our model exhibited consistently the best denoising performance across low to high acceleration rates and excelled in detail restoration, particularly at high acceleration rate scenarios. The adjusted BM3D and SCUNet models showed poor performance in detail preservation, indicated by red arrows (Figure 6) at the 3.5x acceleration scenario. They even suffered mild hallucination, which is demonstrated in 7x acceleration cases. The SCUNet still encountered poor consistency as SNR varies, resulting in severe over-smoothness in highly accelerated scenarios.

**Denoising at low-field MRI as intermediate results**

We conducted a study to translate advanced 3D Magnetic Resonance Fingerprinting[18,19] (MRF) approaches robust at 3T to 0.55T scanners. To boost SNR, we directly employed our denoising method to the five coefficient maps (c1-5) acquired using an optimized FISP-based MRF sequence. The resulting complex-valued, denoised coefficient maps were subsequently fed to fitting algorithms to produce the final parameter maps (T1 and T2). Figure 7 is one example of whole-brain imaging at 1.2mm isotropic resolution, and Supplementary Figure 3 details the denoised results of 5 coefficient maps used for fitting procedures. Employing our denoising approach coefficient maps, we were able to enhance the quality of a 4.5-minute accelerated scan (acceleration factor R=2 in reconstruction), making it superior to that of a standard 9-minute scan, demonstrating our model's capability in enhancing intermediate complex-valued results to facilitate the transfer, improvement, and innovation of advanced MRI technologies at the accessible low-field systems.

# Discussion

This study presents a novel complex-valued imaging transformer deep learning model for universal low-field MRI denoising. Trained on a large-scale and diverse clinical MRI dataset, our model exhibits remarkable generalizability across clinical settings, demonstrating it is a robust tool to enhance image quality and enable faster scans for future development of accessible MRI.

A significant achievement of our model is its robust generalization capability to handle diverse MR image data. Through thorough evaluations covering multiple organs, including internal validation over different noise levels and external validations on different data representations and MRI systems, our model consistently demonstrated the best denoising performance under comprehensive clinical settings. By delivering effective denoising results, our model not only outperforms the non-learning and the SOTA deep learning models but also rivals or even surpasses the specialist model in its specific task setting. This allows our model to efficiently adapt to low-field MRI systems, facilitating faster scans and better imaging quality, even as low-field MRI technology rapidly progresses.

When compared to multi-repetition-averaged scanning, a common technique utilized in low-field systems, our model consistently produces high-quality results from single or fewer repetitions. While averaging repeated scans typically enhances the SNR, this approach often sacrifices scan duration and may encounter diminishing returns due to motion-induced blurring or artifacts. The proportionally increased scan duration not only leads to poor patient experiences as well as low clinical efficiency but also raises the likelihood of positional variations within the scanned regions. These variations may arise from patients' physiologic motions such as breathing, peristalsis, cardiac pulsation, and their difficulty in maintaining stillness over prolonged durations. Supplementary Figure 4 illustrates one knee scan with four repetitions, wherein unexpected motion significantly compromises image quality with blurring and motion artifacts. Our model, however, can leverage single- or fewer-repetition scans to deliver high-quality results that meet or exceed the quality of all the averages.

We also investigated the advantages of using complex-valued image representations throughout the network. We additionally trained our model on magnitude image presentations with the same hyper-parameter sets and noise synthesis method to compare its performance with the default complex-valued model. As shown in Supplementary Figure 5 and Supplementary Table 1, the complex-valued model, learning the rich signal-alike complex representations, consistently outperformed the magnitude one. As the noise level increases, the performance of the magnitude model degrades significantly, resulting in over-smoothness and poor detail preservation when handling low-SNR cases. This demonstrates our model's input of complex-valued images allows for the preservation and utilization of rich MR-specific information that would otherwise be lost with magnitude images. Furthermore, the capability of our model to take in complex-valued input ensures it is a potential asset for subsequent advanced MRI post-processing techniques which could help promote innovation in sequence design, further expanding the application boundaries of low-field MRI technology.

Despite these strengths, it is important to acknowledge the model's inherent limitations. Deep learning models operate within the boundaries of their training data and algorithms, which implies that there is always a non-zero probability of error. One limitation of our model is that although our model allows input images of varying matrix sizes, which is a benefit of our diverse training data and effective resolution augmentation, we cannot guarantee our model's performance in super-high-resolution imaging. This constraint arises due to the scarcity of larger-sized or higher-resolution images (400 * 400 pixels or larger, which are uncommon in MR images without interpolation) in our training dataset. Such limitations necessitate continuous model developments to maintain and improve performance as new data and scenarios emerge. One mitigation is to further curate much more training data from multiple sites for collaborative model training.

## Methods

**Dataset acquisition and pre-processing**

The 3T datasets we used for training and internal validation were sourced from a clinical 3T scanner (Siemens MAGNETOM Vida, UCSF Health). For model development, we gathered 8583 complex-valued images, each comprising 30 slices extracted from 4584 independent raw scans, with 8150 images (95%) randomly allocated for training and 433 images (5%) for model tuning. An additional 150 images from 150 different scans were used for the internal validation.

These images which originated from Siemens raw data were converted to generic ISMRMRD[20] k-space data using the Siemens-to-ismrmrd package[21]. Subsequently, the Gadgetron[22,23] framework was employed to reconstruct the k-space data into complex-valued images and calculate real-valued g-factor maps. Additionally, manual pre-screening was involved to remove phantom and poor-quality scans with aliasing or ultra-low SNR.

For noise synthesizing, we created complex-valued Gaussian noise maps with randomly assigned noise levels (sigma) from 1 to 10. These noise maps were augmented in k-space using k-space filters including resolution reduction, partial Fourier, and Gaussian filters along phase and readout encoding directions. After being transformed back to image space, the augmented complex-valued noise maps were then multiplied by the g-factor maps, producing realistic MRI noise that is in line with the actual signal distribution. These noises were consequently added to the 3T high-SNR images, where each image was first normalized by its mean signal power. This involves adjusting the signal intensity of each image stack to a standard value of 1600 using a factor $k_n$ and then restoring its original signal intensity by the inverse factor $\frac{1}{k_n}$ after the noise was added. This process guarantees uniformity in noise levels throughout the dataset, notwithstanding the substantial variability in intensity values among individual samples.

The 0.55T dataset for external validation was sourced from a clinical-based 0.55T scanner (Siemens MAGNETOM Free. Max, UCSF Health). 150 scans with 4 to 5 repetitions were selected according to the header information stored in the raw data to generate 150 complex-valued images, each composite of all slices from its raw scan. Pre-processing this dataset followed the same reconstruction procedures as the 3T datasets. The low and high SNR pairs were separately generated from the first repetitions and the averages on complex values over all repetitions.

The 0.3T dataset for external validation was collected from the M4Raw test dataset, a publicly available raw MRI image dataset that contains multi-repetitions. The dataset comprises four-channel-coil brain k-space data collected from 25 subjects using a 0.3 Tesla whole-body MRI system (XGY OPER-0.3T). Each subject includes one T1-weighted (6 repetitions), one T2-weighted (6 repetitions), and one FLAIR (4 repetitions) image, totaling 75 images, each containing 18 slices. Single-repetition complex images were generated from the k-space data and combined the coil signals using the root sum of squares over the real and imaginary parts separately. Multi-repetition averaged images were then obtained by calculating the complex average of individual repetitions.

To preserve image intensity and retain information for potential post-processing or conversion to widely used formats like DICOM or NifTI, all datasets in this study were maintained at their original scales. The data taken in by our complex-valued model were kept in the complex64 data type, while their variant used for real-valued models (BM3D, SCUNet, and NAFNet) were obtained by calculating the absolute values, therefore stored as float32 data type.

**Network architecture**

We use the imaging transformer[24] as a foundation architecture in this study, and the overall network structure is illustrated in Figure 8. Instead of directly cutting images into patches and processing them with the standard transformer, the imaging transformer proposed a local (L), global (G), and temporal/slice (T) attention mechanism to utilize different localities for attention computation. This method reduces the total computation and still allows models to capture the local and long-range correlation over different spatial scales and distances. The local spatial attention is computed on neighboring patches within a window, which is directly divided from an image slice or feature map, capturing local patterns within a smaller field of view (FOV). The global attention computes attention over corresponding patches over multiple windows, capturing long-range correlation. The temporal/slice attention computes attention over multiple 2D slices. As the modular design proposed in the imaging attention[24], one set of the G/L/T modules is connected parallelly with batch normalization and mixers into an attention cell[24]. We stacked 2 or 3 attention cells to build an attention block and embedded them into a 2-stage High-Resolution Network[25,26] (HRNet) backbone. This architecture maintains high-resolution representations in the network, which helps achieve better performance in detail preservation.

To handle the significant variation in image intensity (Supplementary Figure 6) over various scan settings for both training and inference, we proposed a pre-network mean power normalization (PowerNorm) mechanism to enhance the model's normalization capacity. Considering the signal characteristics of MRI images, the normalization scales each image stack to a mean signal power of 1600. Compared to the traditional mean-value-based normalization methods or network-inline batch normalization, the pre-network scaling based on mean signal power helps the model receive uniformly distributed data over the entire MRI dataset and, more importantly, mitigate internal intensity shifts within individual images. The mean signal power of one image stack $p_n$ is defined as:

$$p_n = \frac{1}{N}\sum_{i}^{N}|x_i|^2$$

The scaling factor $k_n$ is defined as:

$$k_n = \sqrt{\frac{P}{p_n}}$$

Here $x_i$ denotes the complex value at voxel $i$ and $N$ is the total number of voxels in image stack $n$. $P$ is the mean signal power with a value set to 1600. Before the network, the input image stack is multiplied by $k_n$ for normalization. After the network, the inference result is restored to its original intensity level by the factor $\frac{1}{k_n}$.

Charbonnier loss[27] and perceptual loss[28] were used in training to mitigate over-smoothness and promote visually desirable results. A separate or combination of the two losses has been proven robust and effective in different denoising and deblurring tasks over both natural images or medical images[29–32]. In this study, we used the weighted sum between the two losses as the final loss function to optimize the proposed network. Given the ground-truth and denoised image are $\hat{I}$ and $I'$. $\epsilon = 10^{-3}$ is a constant in all the experiments. Charbonnier loss is defined by:

$$\ell_c(I', \hat{I}) = \sqrt{\|I' - \hat{I}\|^2 + \epsilon^2},$$

Let $\phi_j(|I'|)$ and $\phi_j(|\hat{I}|)$ denote the feature maps extracted from the $j^{th}$ layer of a network when processing $I'$ and $\hat{I}$, respectively. This study uses the first 23 layers of a pre-trained VGG16 network for feature extraction. $C_j$, $H_j$, and $W_j$ are the dimensions of the $j^{th}$ layer feature maps: channels, height, and width. Perceptual loss is then defined by:

$$\ell_p(I', \hat{I}) = \frac{1}{C_j H_j W_j} \|\phi_j(|I'|) - \phi_j(|\hat{I}|)\|_2^2,$$

The final combined loss $\ell$ is defined by:

$$\ell(I', \hat{I}) = \ell_c(I', \hat{I}) + 0.1\ell_p(I', \hat{I})$$

**Training and experimental settings**

During training, we performed data augmentation on the training dataset with random flips, overall intensity perturbation from 0.3 to 3, and k-space matrix resizing with a ratio from 0.5 to 1.5. To balance the model performance, robustness, and training time consumption, we performed multi-size patch training and set the spatial patch size to 32 × 32 pixels × 30 slices and 64 × 64 pixels × 30 slices. The global batch size was set to 32. To accelerate training on a large dataset, the network is optimized by a Sophia optimizer[33] ($\beta_1 = 0.9, \beta_2 = 0.95$) with an initial learning rate of $1e^{-4}$ and a weight decay of 0.1. The model was trained on 4 A6000 (48G) GPUs with 75 epochs, and the checkpoint with the best validation score was selected as the best model. A single GPU was used for inference, and it took 4 seconds to generate a denoised image stack with a typical shape of 256 × 256 pixels × 30 slices. During performance evaluation, an adjusted BM3D incorporating a wavelet-based dynamic estimation of noise standard deviation[34] was proposed in this study to ensure consistent denoise performance when handling datasets of such significant variance. The sigma parameter for BM3D was set to 1.15 times the noise standard deviation estimation on the middle slice of each image stack.

**Evaluation metrics**

To quantitively assess the quality of denoising performance, we employed several widely recognized evaluation metrics: PSNR, SSIM, and NRMSE. PSNR is a standard measure that quantifies the ratio between the maximum possible power of a signal and the power of corrupting noise that affects the fidelity of its representation. It is especially effective for assessing the quality of reconstructed high-SNR images. A higher PSNR value typically indicates better denoising performance. SSIM measures the similarity between two images by comparing local patterns of pixel intensities that have been normalized for luminance and contrast. NRMSE provides a normalized measure of deviations between the denoised image and the ground truth, scaling the root mean square error by the range of possible pixel values.

Since all baseline models operate on real-valued images, to ensure a fair comparison, we converted both the complex-valued outputs of our model and the ground-truth/reference images to their absolute values before computing the metrics. For PSNR and SSIM calculations, the scores for each pair of images were calculated as the mean scores over all individual slices within the stack. Considering a wide range of image intensity values, we used a dynamic data range for each testing pair, which is set from the minimum value to the maximum value of that pair.

**Qualitative evaluation by radiologists**

In addition to quantitative metrics, two board-certified radiologists conducted a blind qualitative assessment of the denoising results. This evaluation aimed to verify the model's clinical applicability, emphasizing its potential to meet the demands of medical imaging in real-world diagnostic settings. Each radiologist independently reviewed randomly ordered images acquired through three approaches over the 0.55T validation set, including 150 single repetition images, 150 multi-repetition averaged ones, and 150 denoised using our model. The totaling 450 images were directly converted from the complex-valued image format to the standard DICOM images (scaled to a 0-8192 range, 16 bit), ensuring high precision fidelity to the original data and facilitating assessment with routine DICOM reading tools used in clinical practice.

Using a grading scale from 1 (poor quality) to 5 (excellent quality), the radiologists scored each image on four criteria: noise suppression, sharpness, detail preservation, and overall quality, which are detailed in Supplementary Table 2. This grading system provided insights beyond quantitative metrics, focusing on aspects of clinically relevant image utility and diagnostic value. The gradings sought to capture the practical implications of denoising, such as over-smoothing or the loss of essential details, as well as the consistency of noise removal across varied clinical images.

**Statistical analysis**

Comparative statistical analysis of model performance was conducted using paired t-tests in Supplementary Table 3-5, with a P-value of less than 0.05 indicating statistical significance. For analyzing radiologists' qualitative evaluation, a Bland-Altman plot was used to assess the agreement between the two graders in scoring overall image quality in Supplementary Figure 8, and the Intraclass correlation coefficient (ICC) of 0.59 is considered moderate.

## Dataset availability

We provide access to web links for public data used in our study. The M4Raw dataset used in this study is the M4Raw multicoil test subset and can be accessed at https://zenodo.org/records/8056074. Since there are extremely limited open-source raw MRI datasets containing multi-averages during clinical scans, we open-access our 0.55T validation set for potential usage as a training or validation source with real high- and low-SNR pairs.

## Code availability

The training script, inference script, and trained model weights have been publicly available at https://github.com/zherenz/ImT-MRD. The codebase for the imaging transformer architecture is available at https://github.com/NHLBI-MR/FMImaging.git.

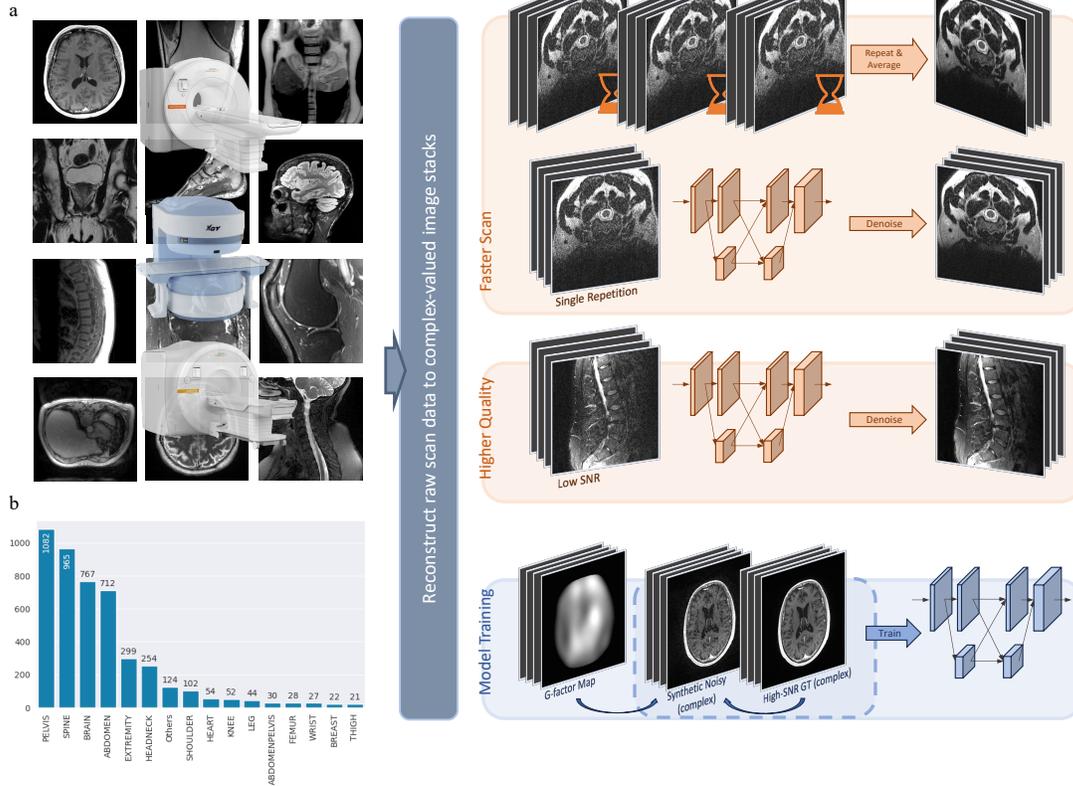

**Fig.1 Data acquisition and network pipeline. a** An overview of the training and inference pipeline: model training includes 8583 diverse complex-valued high-SNR image stacks reconstructed from 3T raw scan data with their noisy variants synthesized based on g-factor maps. In the inference stage, the denoise network is utilized to accelerate clinical examination or improve image quality by taking single-repetition or low-SNR complex-valued images from low-field scanners after reconstruction. **b** The number of raw MRI scans in anatomic regions used for model development.

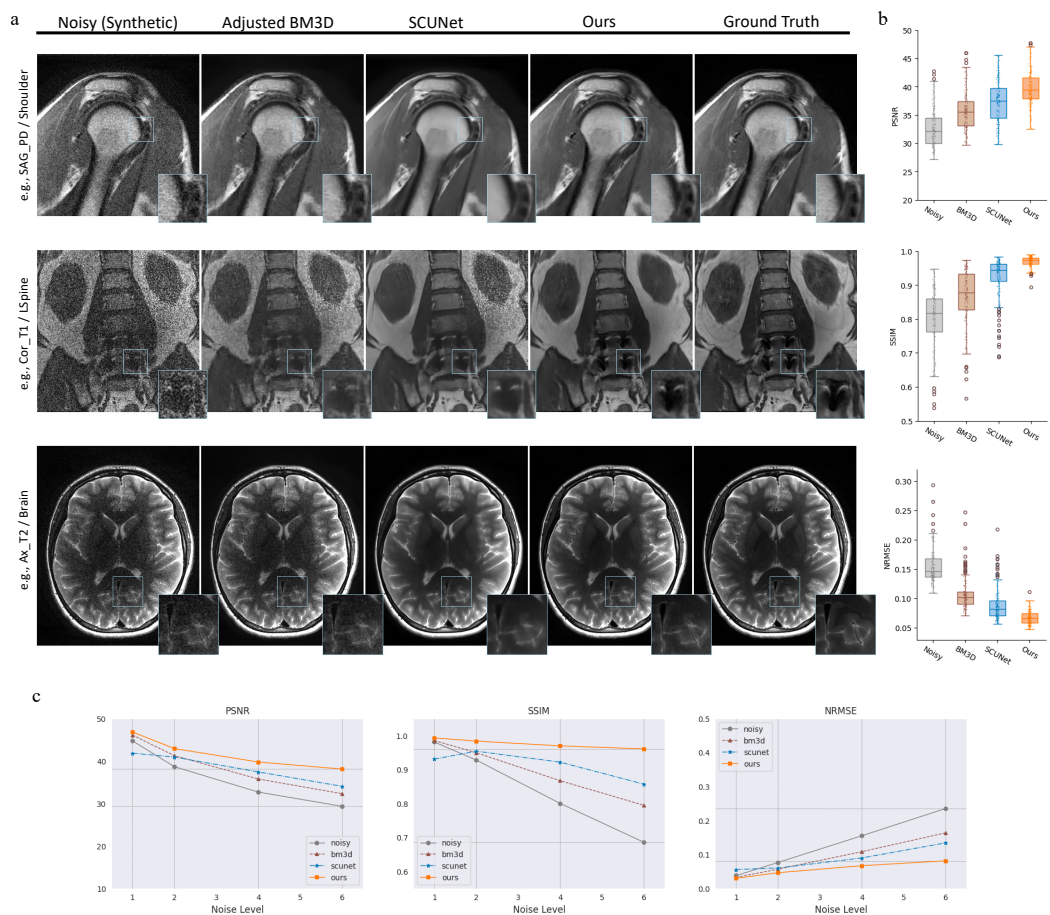

**Fig.2 Evaluation results on the internal validation set with varying noise levels. a** Visualized examples on the internal validation set with level-4 noise synthetic from a 3T scanner. The three examples are shoulder, L-spine, and brain scans in sagittal proton-density-weighted (PD), coronal T1-weighted, and axial T2-weighted sequence contrasts. **b** Performance distribution of 150 validation cases at noise level 4 in terms of PSNR, SSIM, and NRMSE scores. The center line within the box represents the median value, with the bottom and top bounds indicating the 25th and 75th percentiles. Whiskers are showing the 1.5 of the interquartile range. Scatter dots corresponding to data points are plotted individually with those beyond the whiskers considered outliers. **c** The average performance scores over the internal validation set at noise levels 1, 2, 4, and 6.

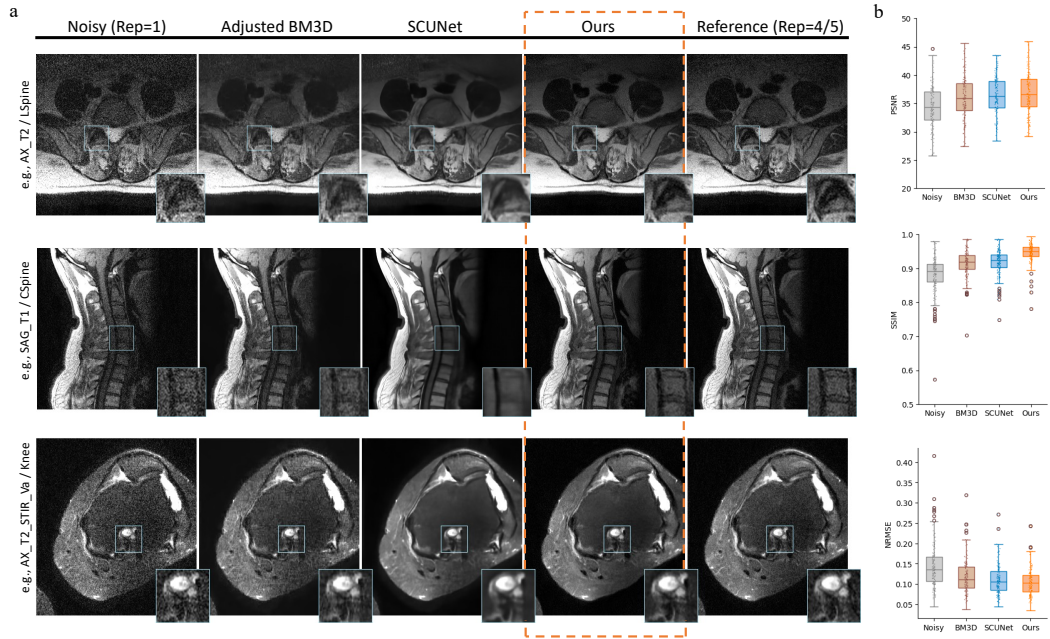

**Fig.3 Evaluation results on the 0.55T external validation set. a** Visualized examples on the external validation set with diverse scan protocols from a 0.55T scanner. The three examples are L-spine, C-spine, and knee scans in axial T2-weighted, sagittal T1-weighted, and axial T2-weighted Short Tau Inversion Recovery (STIR) with Variable Flip Angle (VA) sequence contrasts. **b** Performance distribution of 150 validation cases in terms of PSNR, SSIM, and NRMSE scores. The center line within the box represents the median value, with the bottom and top bounds indicating the 25th and 75th percentiles. Whiskers are showing the 1.5 of the interquartile range. Scatter dots corresponding to data points are plotted individually with those beyond the whiskers considered outliers.

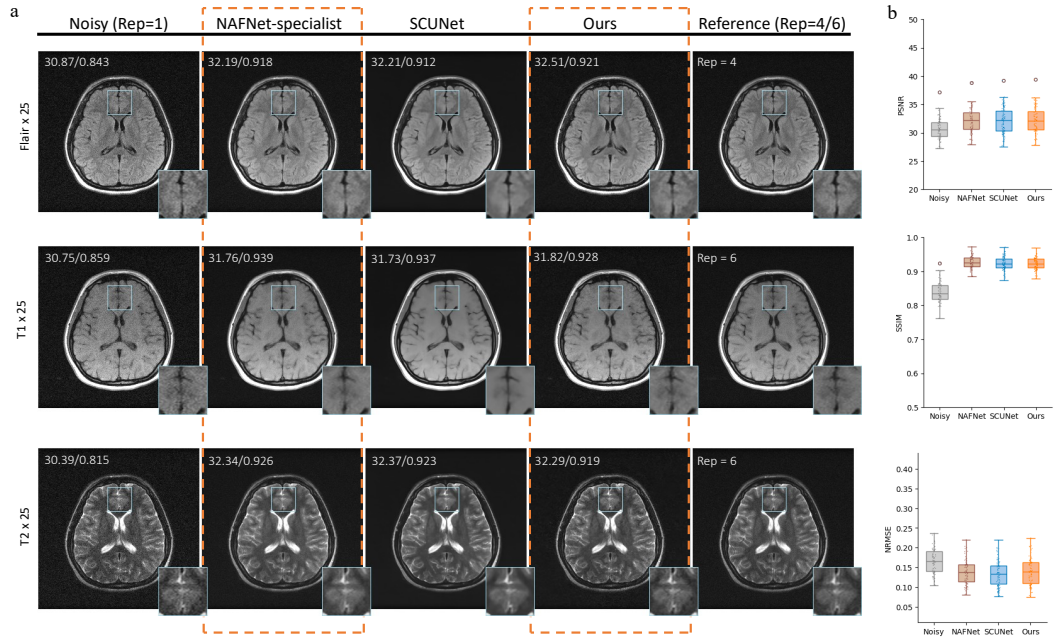

**Fig.5 Evaluation results on the 0.3T external validation set. a** Visualized examples on the external validation set of brain scans from a 0.3T scanner. The three examples are a set of brain scans in Fluid Attenuated Inversion Recovery (Flair), T1, and T2 contrasts. Mean PSNR and SSIM scores were calculated for each contrast group and displayed on the top-left of the figures. **b** Performance distribution of 75 validation cases in terms of PSNR, SSIM, and NRMSE scores. The center line within the box represents the median value, with the bottom and top bounds indicating the 25th and 75th percentiles. Whiskers are showing the 1.5 of the interquartile range. Scatter dots corresponding to data points are plotted individually with those beyond the whiskers considered outliers.

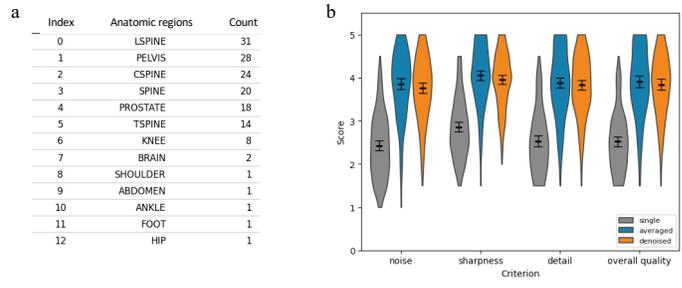

**Fig.4 Qualitative evaluation results on the 0.55T external validation set. a** Distribution of anatomic regions included in the dataset. **b** Performance distribution of the dataset in terms of two radiologists' average grading scores on noise level, sharpness, detail preservation, and overall image quality criteria, where higher scores indicate better performance. The width of each violin represents the density of the data, with a center line indicating the median score for each group. Error bars are included to depict the 95% confidence intervals around the median values.

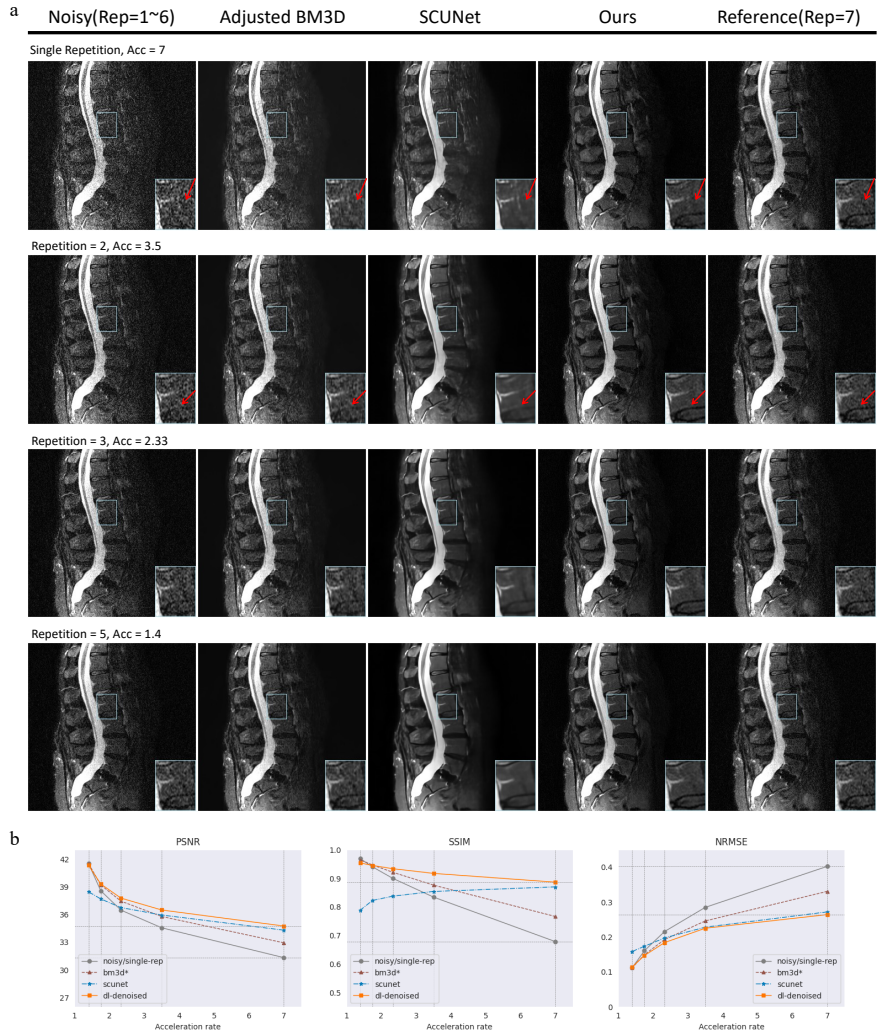

**Fig.6 Evaluation of denoising performance under 1.4-7x acceleration. a** Visualized examples using the averages of 1, 2, 3, and 5 repetitions as the input for denoising, corresponding to acceleration rates of 7, 3.5, 2.33, and 1.4. The average of 7 single-repetitions is utilized as a non-accelerated reference visualized in the last column. **b** Performance scores under different acceleration rates.

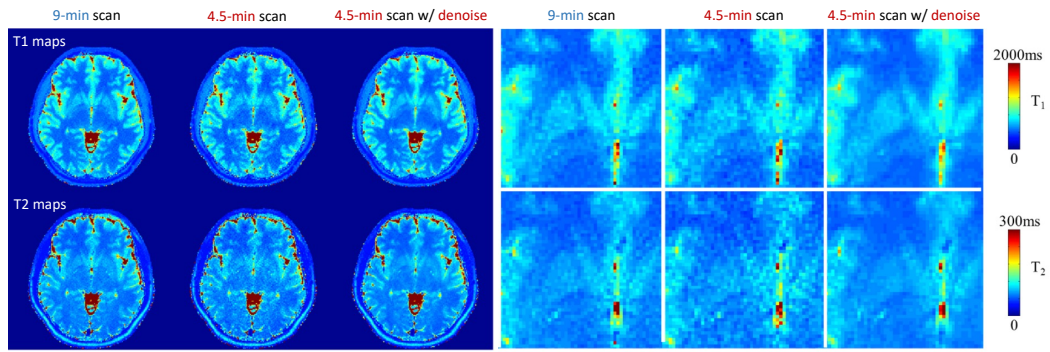

**Fig.7 Boost low-field MRF with denoised coefficient maps.** Visualized a brain MRF scan example with a 9-min acquisition time. Based on five acquired coefficient maps, non-accelerated T1 and T2 maps were reconstructed by using measured trajectory, shown in Column 1. An acceleration factor R=2 was implemented with the results visualized in Column 2, as well as results involving additional denoising applied to the coefficient maps in Column 3.

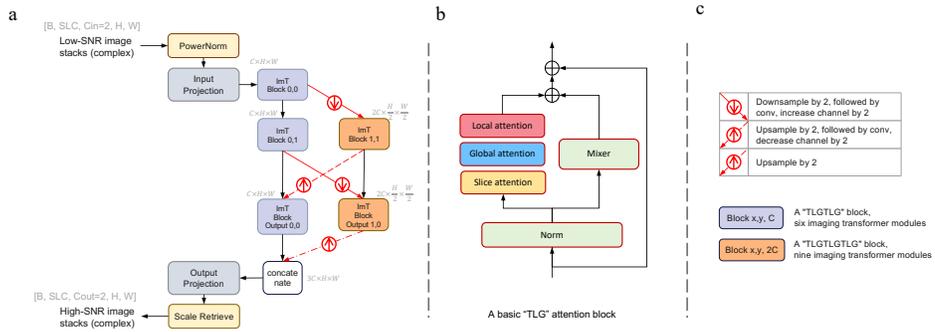

**Fig.8 Architecture of the denoising network. a** Pipeline overview: the network takes multi-slices complex-valued images (real and imaginary parts) as two-channel inputs. It outputs multi-slices complex images in two channels. **b** Illustration of a basic spatial-temporal imaging transformer (ImT) block includes one temporal, one global, and one local attention (TLG) unit. **c** Legends.

# Supplementary

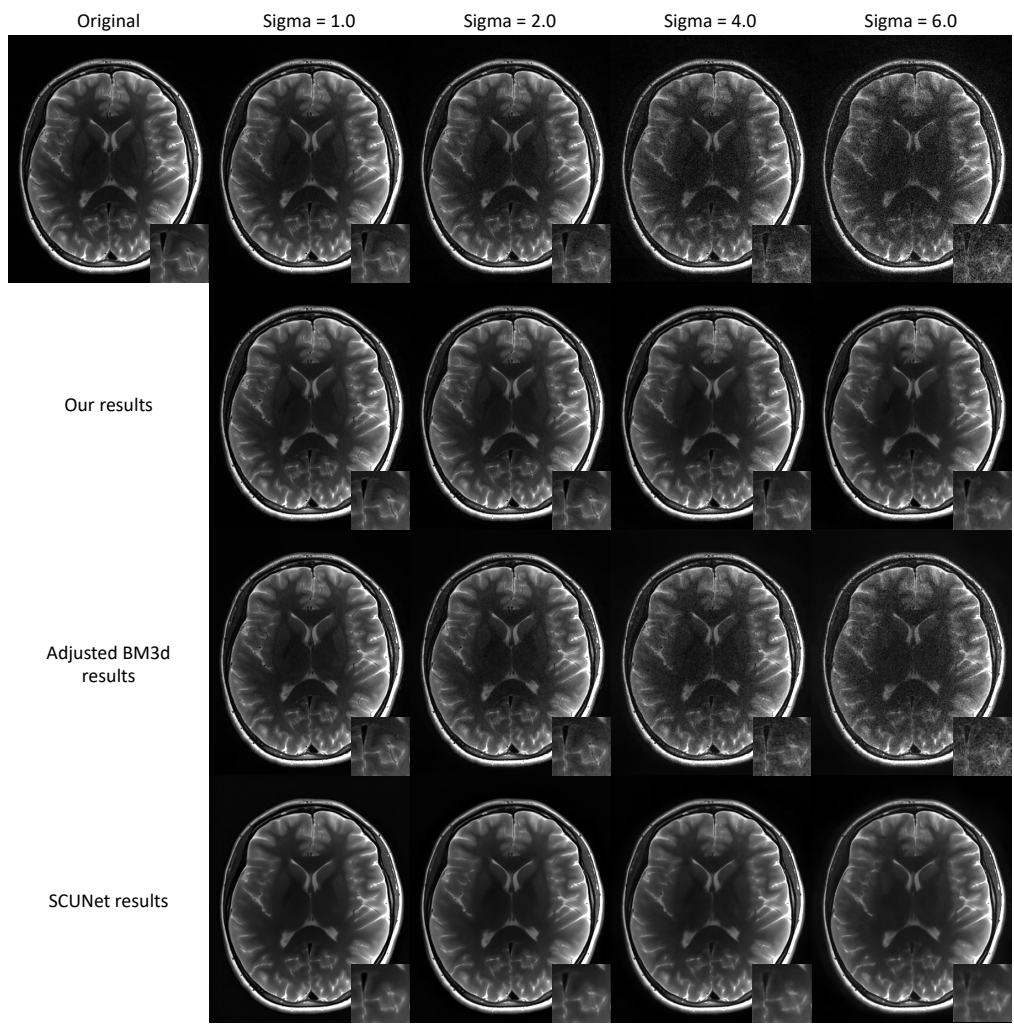

**Supplementary Fig.1 Denoise results regarding different noise levels.** The 1st row visualizes the ground truth and noisy images with noise levels 1, 2, 4, and 6. 2nd to 4th row visualizes denoising results from different models.

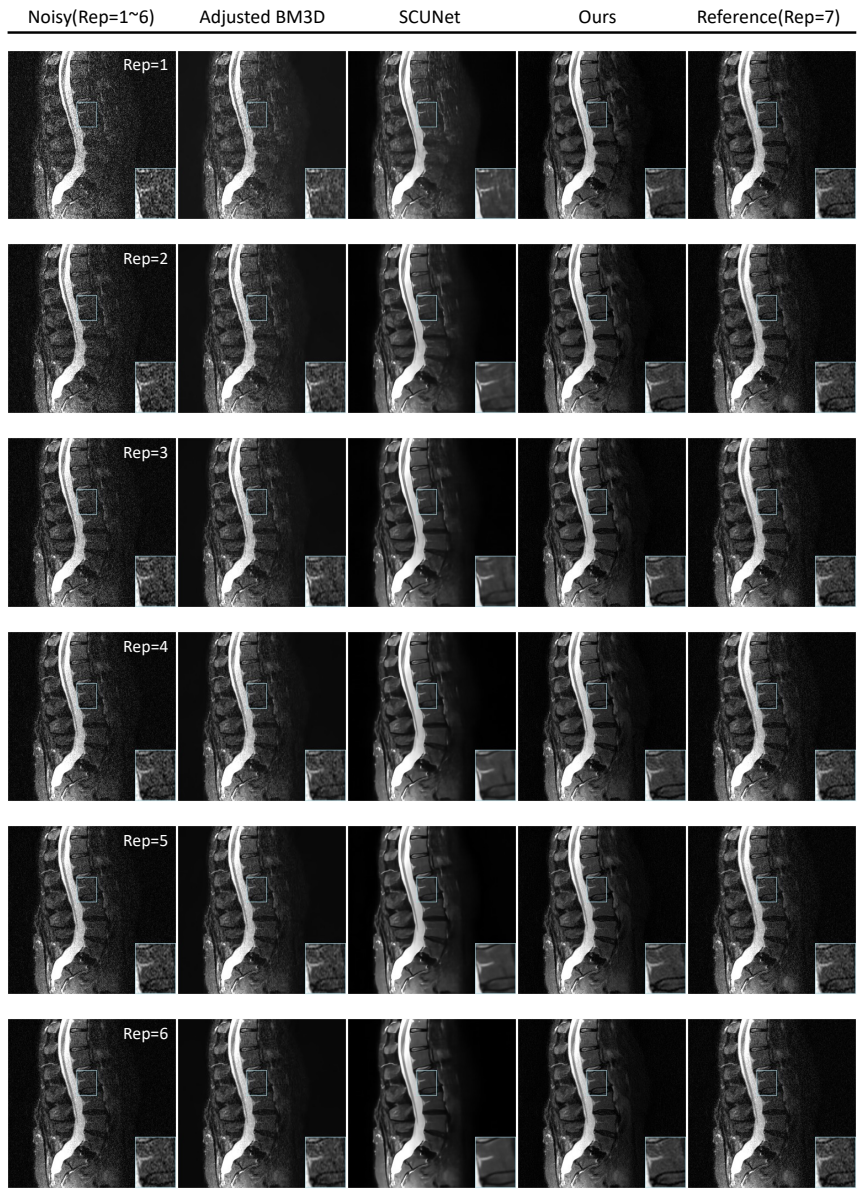

**Supplementary Fig.2 Denoise results with the input image of 1 to 6 repetitions.** Visualized examples using the averages of 1, 2, 3, 4, 5, and 6 repetitions as the input for denoising, corresponding to acceleration rates of 7, 3.5, 2.33, 1.75, 1.4, and 1.17. The average of 7 single-repetitions is utilized as a non-accelerated reference visualized in the last column.

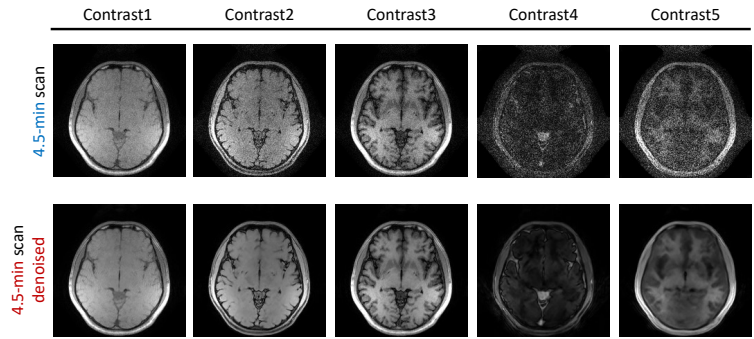

**Supplementary Fig.3 Visualization of original MRF coefficient maps and the denoised ones.** Our denoising is applied to each coefficient map (contrast 1 to 5, complex-valued), separately. The denoising results were fed to the following MRF fitting algorithms.

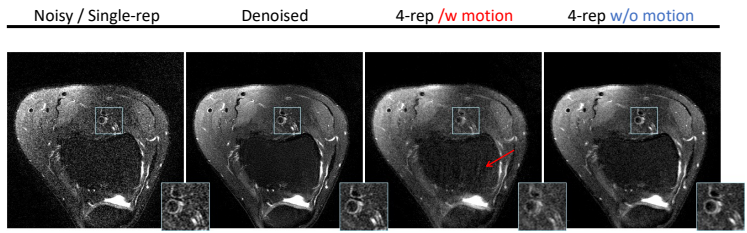

**Supplementary Fig.4 Potential motion artifacts derived from multiple-repeated scans.** The visualized example w/ and w/o slight motion during a 4-times repeated scan and a denoise result by our model from a single repetition. The red arrow and detail view in the 3rd column indicate the motion artifacts and blur derived from the multiple-repeated scan.

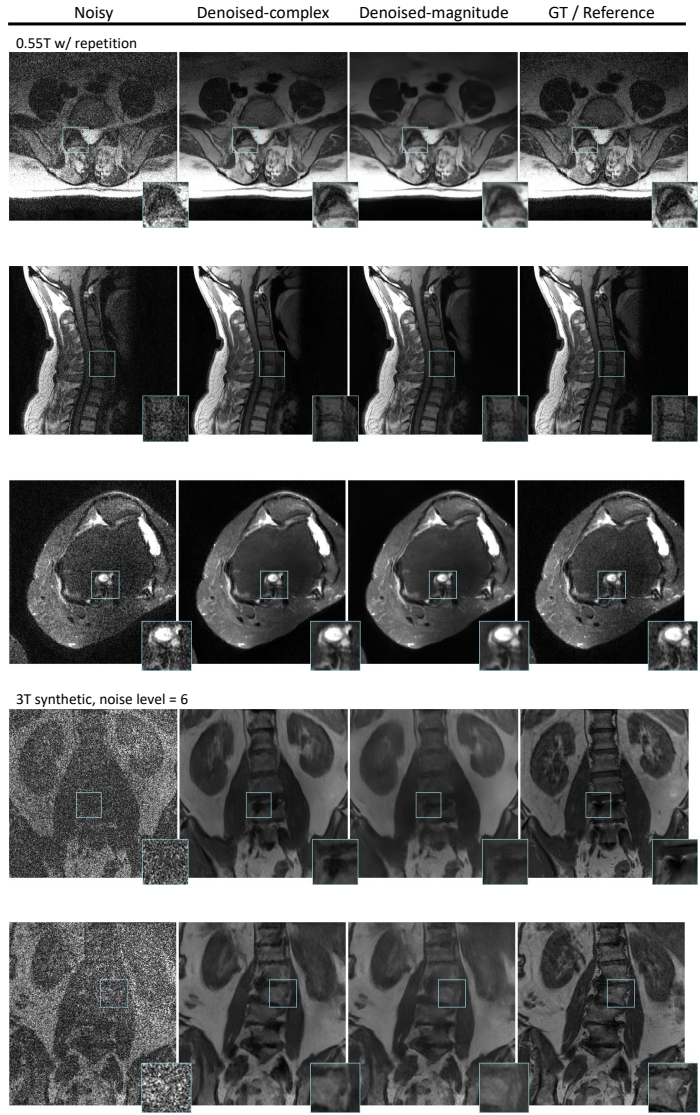

**Supplementary Fig.5** Examples of denoising results from complex and magnitude models.

| Metric | Noise Level | Noisy Input | Model-magnitude | Model-complex |
|---|---|---|---|---|
| PSNR | 1 | 44.73 ± 3.60 | 46.72 ± 3.31 | **46.81 ± 3.28** |
| | 2 | 38.65 ± 3.56 | 42.59 ± 3.14 | **42.92 ± 3.07** |
| | 4 | 32.67 ± 3.38 | 39.26 ± 3.02 | **39.76 ± 3.02** |
| | 6 | 29.33 ± 3.23 | 37.40 ± 3.07 | **38.10 ± 3.10** |
| SSIM (x1E-2) | 1 | 98.08 ± 1.45 | 99.24 ± 0.44 | **99.30 ± 0.35** |
| | 2 | 92.77 ± 4.49 | 97.95 ± 1.20 | **98.35 ± 0.79** |
| | 4 | 80.02 ± 8.72 | 95.91 ± 2.33 | **96.95 ± 1.53** |
| | 6 | 68.58 ± 10.61 | 94.08 ± 3.48 | **96.06 ± 1.86** |
| NRMSE (x1E-2) | 1 | 3.78 ± 0.70 | 2.99 ± 0.48 | **2.96 ± 0.48** |
| | 2 | 7.64 ± 1.41 | 4.81 ± 0.76 | **4.63 ± 0.69** |
| | 4 | 15.50 ± 2.87 | 7.11 ± 1.40 | **6.69 ± 1.12** |
| | 6 | 23.52 ± 4.44 | 8.84 ± 2.07 | **8.12 ± 1.46** |

**Supplementary Table 1** Performance scores of complex and magnitude models over the internal validation set on noise levels 1 to 6. Scores are displayed as mean values and standard deviations.

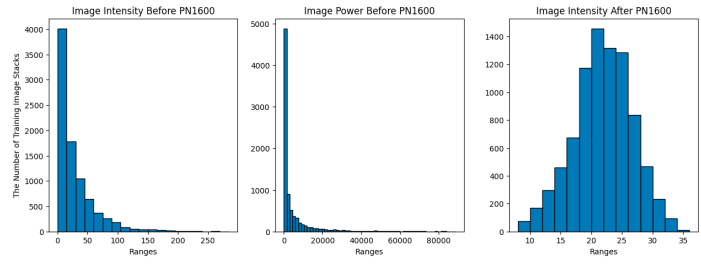

**Supplementary Fig.6** The mean image intensity and power distribution over the training dataset before and after the power norm.

| Score | Noise intensity | Image sharpness | Detail preservation | Overall quality |
|---|---|---|---|---|
| 5 | very little to no noise | excellent sharpness | all details are well-preserved | excellent |
| 4 | minimal noise | very sharp, minimal blurriness | most details are retained | good |
| 3 | some noticeable noise | moderately sharp, noticeable blurriness in some areas | most major details are clear, some details may be lost | average |
| 2 | a significant amount of noise | blurred | significant loss of details | poor |
| 1 | excessive noise | severely blurred | almost no clear details | terrible |

**Supplementary Table 2** Criterion for image quality grading by radiologists.

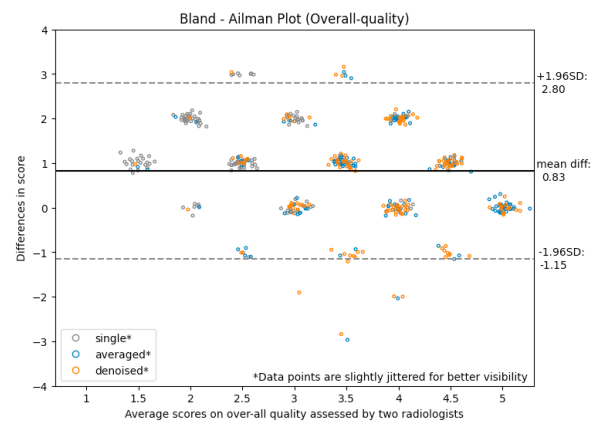

**Supplementary Figure 7** Bland-Altman plot assessing the agreement between the two graders in scoring overall image quality.

| Metric | Noise Level | Noisy Input | Adjusted BM3D | SCUNet | Ours |
|---|---|---|---|---|---|
| PSNR | 1 | 44.73 ± 3.60^ | 46.10 ± 3.43^ | 41.84 ± **2.16**^ | **46.81** ± 3.28 |
|  | 2 | 38.65 ± 3.56^ | 41.24 ± 3.37^ | 40.92 ± **2.46**^ | **42.92** ± 3.07 |
|  | 4 | 32.67 ± 3.38^ | 35.76 ± 3.35^ | 37.42 ± 3.64^ | **39.76 ± 3.02** |
|  | 6 | 29.33 ± 3.23^ | 32.30 ± 3.33^ | 34.04 ± 4.03^ | **38.10 ± 3.10** |
| SSIM (x1E-2) | 1 | 98.08 ± 1.45^ | 98.51 ± 1.16^ | 93.05 ± 6.45^ | **99.30 ± 0.35** |
|  | 2 | 92.77 ± 4.49^ | 94.92 ± 3.84^ | 95.38 ± 3.92^ | **98.35 ± 0.79** |
|  | 4 | 80.02 ± 8.72^ | 86.70 ± 8.20^ | 92.16 ± 6.27^ | **96.95 ± 1.53** |
|  | 6 | 68.58 ± 10.61^ | 79.46 ± 10.76^ | 85.68 ± 9.69^ | **96.06 ± 1.86** |
| NRMSE (x1E-2) | 1 | 3.78 ± 0.70^ | 3.25 ± 0.72^ | 5.54 ± 2.27^ | **2.96 ± 0.48** |
|  | 2 | 7.64 ± 1.41^ | 5.67 ± 1.23^ | 6.02 ± 1.99^ | **4.63 ± 0.69** |
|  | 4 | 15.50 ± 2.87^ | 10.81 ± 2.70^ | 8.96 ± 2.75^ | **6.69 ± 1.12** |
|  | 6 | 23.52 ± 4.44^ | 16.36 ± 4.29^ | 13.41 ± 4.47^ | **8.12 ± 1.46** |

**Supplementary Table 3** Internal validation results of original noisy input, adjusted BM3D, SCUNet, and our model on different noise levels in terms of PSNR, SSIM, and NRMSE. Scores are displayed as mean values with standard deviation. The paired t-test was performed for the validation results of each target. Scores marked with ^ denote statistical significance between the result of our model and that of the compared method (p-value < 0.05).

| Metric | Noisy (Rep=1) | Adjusted BM3D | SCUNet | Ours |
|---|---|---|---|---|
| PSNR | 34.41 ± 3.71^ | 35.98 ± 3.67^ | 36.39 ± **3.41**^ | **36.80** ± 3.61 |
| SSIM (x1E-2) | 88.26 ± 5.47^ | 91.47 ± 3.79^ | 91.87 ± 3.61^ | **94.62 ± 2.80** |
| NRMSE (x1E-2) | 14.39 ± 5.48^ | 11.82 ± 4.27^ | 11.05 ± 3.65^ | **10.62 ± 3.48** |

**Supplementary Table 4** 0.55T external validation results of single-repeated noisy input, adjusted BM3D, SCUNet, and our model compared to the multi-repeated references in terms of PSNR, SSIM, and NRMSE. Scores are displayed as mean values with standard deviation. The paired t-test was performed for the validation results of each target. Scores marked with ^ denote statistical significance between the result of our model and that of the compared method (p-value < 0.05).

| Metric | Noisy (Rep=1) | NAFNet-specialist | SCUNet | Ours |
|---|---|---|---|---|
| PSNR | 30.67 ± 1.77^ | 32.10 ± **2.07**^ | 32.10 ± 2.31^ | **32.21** ± 2.23 |
| SSIM (x1E-2) | 83.93 ± 2.90^ | **92.73** ± 1.74^ | 92.41 ± 1.94 | 92.25 ± **1.67** |
| NRMSE (x1E-2) | 16.53 ± 3.05^ | **13.80** ± 3.10 | 13.39 ± 3.24^ | 13.89 ± 3.34 |

**Supplementary Table 5** 0.3T brain external validation results of single-repeated noisy input, specialist NAFNet, SCUNet, and our model compared to the multi-repeated references in terms of PSNR, SSIM, and NRMSE. Scores are displayed as mean values with standard deviation. The paired t-test was performed for the validation results of each target. Scores marked with ^ denote statistical significance between the result of our model and that of the compared method (p-value < 0.05).